\documentclass[10pt, journal, compsoc]{IEEEtran}

\usepackage{amsmath}
\usepackage{amssymb}
\usepackage[nocompress]{cite}   
\usepackage{graphicx}

\begin{document}

\title{A Simple and Realistic Pedestrian Model\\for Crowd Simulation and Application}

\author{Wonho~Kang~and~Youngnam~Han
\IEEEcompsocitemizethanks{\IEEEcompsocthanksitem W.~Kang and Y.~Han
are with the Department of Electrical Engineering, Korea Advanced
Institute of Science and Technology (KAIST), Daejeon,
Korea, e-mail: \{wonhoz, ynhan\}@kaist.ac.kr.}
\thanks{Manuscript written August 9, 2014; released August 9, 2017.}}





\IEEEtitleabstractindextext{

\begin{abstract}
The simulation of pedestrian crowd that reflects reality is a major
challenge for researches. Several crowd simulation models have been
proposed such as cellular automata model, agent-based model, fluid
dynamic model, etc. It is important to note that agent-based model
is able, over others approaches, to provide a natural description of
the system and then to capture complex human behaviors.
%

In this paper, we propose a multi-agent simulation model
%
%
in which pedestrian positions are updated at discrete time
intervals.
%
It takes into account the major normal conditions of a simple
pedestrian situated in a crowd such as preferences, realistic
perception of environment, etc.
%
Our objective is to simulate the pedestrian crowd realistically
towards a simulation of believable pedestrian behaviors.
%
Typical pedestrian phenomena, including the unidirectional and
bidirectional movement in a corridor as well as the flow through
bottleneck, are simulated.
%
The conducted simulations show that our model is able to produce
realistic pedestrian behaviors.
%
The obtained fundamental diagram and flow rate at bottleneck agree
very well with classic conclusions and empirical study results.
%
It is hoped that the idea of this study may be helpful in promoting
the modeling and simulation
%
%
of pedestrian crowd in a simple way.
\end{abstract}

} 

\maketitle 

\IEEEdisplaynontitleabstractindextext
\IEEEpeerreviewmaketitle

\IEEEraisesectionheading{\section{Introduction}\label{sec:I}} 
\IEEEPARstart{P}{edestrian}
crowd is a phenomenon that can be observed in several situations
such as in the street, intersection, square, etc. A number of
researchers have been interested in studying this phenomenon.
%
In fact, when the density of the crowd is high, several accidents
and disasters could occur. More the crowd becomes dense, more the
situation is dangerous.
%
Therefore, researchers face an increasing challenge to find
solutions that seriously improve pedestrian management and safety
during crowd. In such a context, simulation is an appropriate tool.
It could be really interesting to be able to simulate pedestrian
movements in such environments.

For this purpose, several crowd simulation models have been
proposed. These models are generally classified into two categories
\cite{ref:conf1}:
macroscopic models that include regression models and flow dynamic
models, and microscopic models including cellular automata models
and agent-based models.
%
Each model can be specifically categorized into continuous,
discrete, and semicontinuous according to the space and time of the
system.
%
The macroscopic approaches simulate the behavior of the crowd as a
whole and do not consider individual features. However, the
microscopic approaches are interested in the behaviors, actions, and
decisions of each pedestrian as well as interactions with others
\cite{ref:Conf1-2}. Therefore, the microscopic models allow us to
obtain more realistic results of simulation. This is why we will
focus only on microscopic models and more specifically on
agent-based models. In fact, the agent-based models are able, over
others approaches, to be flexible, to provide a natural description
of the system, and to capture complex human behaviors
\cite{ref:Conf1-3}.

In this paper, a semicontinuous pedestrian model is developed
%
%
in which the space occupied by pedestrians is continuously evolving,
but time is measured by intervals.
%
It is worth noting that each pedestrian in the proposed model is
regarded as a self-adapted agent, and the movement is implemented in
a utility maximization approach by considering various human
factors.
%
The model is able to calculate normal pedestrian distributions in
space and pedestrian movements at normal step frequencies over time.
%
The goal of our work is to model realistic pedestrian crowd by
simulating realistic pedestrian behaviors. Face this challenge; we
need to carefully study human nature such as preferences, realistic
perception of environment, etc.
%
Rules governing the selection of step size and
moving direction guarantee that the model can be used to accurately
compute the distance and speed of pedestrian movement. In the model,
pedestrians are not treated as particles, and the sizes of their
bodies are considered. The model is thus suitable for simulating the
movements of dense crowds.

The rest of our paper is as follows: section~\ref{sec:II} describes
related works. In section~\ref{sec:III}, we propose our
system model which takes into account the various human nature
factors. Section~\ref{sec:IV} presents our simulations and a
discussion of the work. Finally, we conclude the paper and give
perspectives in section~\ref{sec:V}.

%
\section{Related Works}\label{sec:II}
There are several crowd simulation models which can be classified
into two categories: macroscopic models and microscopic models.
Macroscopic approaches include regression models \cite{ref:Conf1-6}
and flow dynamic models \cite{ref:Conf1-7, ref:Conf1-8}. Microscopic
approaches include cellular automata models
\cite{ref:SCI2-24, ref:Conf1-17, ref:Conf3-3}
and agent-based models
\cite{ref:Conf1-3, ref:agent2, ref:Conf1-28, ref:SCI1-31,
ref:SCI1-32, ref:SCI1-34}.
The macroscopic approaches simulate the behavior of the crowd as a
whole. In fact, macroscopic models do not consider individual
features such as physical abilities, direction of movement, and
individual positioning \cite{ref:Conf1-9}. This causes a lack of
realism. On the other hand, the microscopic approaches are
interested in the behaviors, actions, and decisions of each
pedestrian as well as interactions with others \cite{ref:Conf1-2}.
Therefore, the microscopic models allow us to obtain more realistic
outputs of simulation. For this reason, in our work, we adopt a
microscopic approach.
%
Among the specific categories of the microscopic systems, we focus
on semicontinuous multi-agent model
%
%
in which pedestrian space is continuous, and pedestrian positions
are updated at discrete time intervals.
%
In the following, we describe the major microscopic models.

\subsection{Cellular Automata Model}\label{sec:II_A}
A cellular automaton
%
%
\cite{ref:SCI2-24, ref:Conf1-17, ref:Conf3-3}
is a collection of cells on a grid. Each cell contains a state
chosen among a finite set and can change over time. The pedestrian
transitions from one state to another consist firstly on the
awareness of the environment and secondly on the possibility to move
to the state of its neighboring cells. Finally, the transitions of
states are executed.
%
The change of cell state is based on a set of rules that are applied
simultaneously to all grid cells, producing a new generation of
cells which depends entirely on the previous generation.

Cellular automata model is a simple and fast implementation, but
criticized for lacking realism.
In fact, this model limits the pedestrian spatial movement like a
checkerboard.
%
%
Another limitation of cellular automata is the difficulty to model
changing pedestrian velocities and to simulate heterogeneous
pedestrian behaviors.

%
%
\begin{figure}[!t]
\centering
\includegraphics[width=0.4\textwidth]{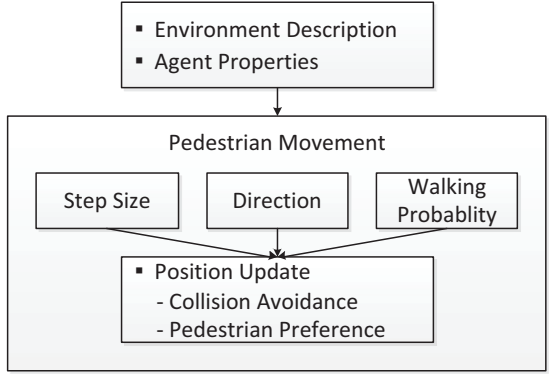}
\caption{Architecture of system model in overview.}
\label{fig:systemmodel}
\end{figure}
\section{System Model}\label{sec:III}
Our system is based on a semicontinuous microscopic approach and
particularly in a multi-agent architecture,
%
%
in which each pedestrian moves a certain step size in the moving
direction at each discrete time instant.
%
The overview of system architecture is displayed in
Fig.~\ref{fig:systemmodel}. The description on the environment and
the properties of the agents are required as inputs for the system.
%
The pedestrian movement is represented as step size, direction, and
walking probability, which are used to update the position of a
pedestrian.
%
In the model, a rule of determining step size and movement direction
has been used to avoid pedestrian overlapping each other,
%
and the design of the pedestrian behavior is based on
%
%
the perception of the environment.
%
%
This combination has allowed us to obtain a simple and realistic
pedestrian model for crowd simulation. In our model, each pedestrian
is represented by an agent,
%
who is required to move step by step toward the goal while adjusting
the step size and moving direction according to the environment and
avoiding collisions with other pedestrians and obstacles.
%
%
The $i^{\rm{th}}$ pedestrian agent at time $t$ is characterized by:

\begin{itemize}
\item Step size: The actual step size of an agent is defined as
$\alpha {l_{i,t}}$ with step scale factor $\alpha$ and desired step
size of ${l_{i,t}} \sim \mathcal{N}\left( {{\mu _{l,i}},\sigma
_{l,i}^2} \right)$.
\item Direction: The possible moving direction is described by
${{\theta _{i,t}} + \phi }$ with direction shift factor $\phi $ and
desired direction of ${\theta _{i,t}} \sim \mathcal{N}\left( {{\mu
_{\theta ,i}},\sigma _{\theta ,i}^2} \right)$.
\item Position: The environment of our system is indexed by a
Cartesian coordinate system as $s \!=\! \left( {x,y} \right), \; x,y
\in \mathbb{R}$. Thus, a position is defined as the pair of
coordinates ${s_{i,t}} \!=\! \left( {{x_{i,t}},{y_{i,t}}} \right)$
in the Cartesian coordinate system.
\end{itemize}
%
Each agent has its own autonomous behavior. Indeed, the behavior of
a pedestrian agent is divided into three phases: strategic phase,
tactical phase, and operational phase.

\subsection{Strategic Phase}\label{sec:III_A}
%
At each
time step $t$, we consider that a pedestrian agent moves through a
given environment, which represents for street, intersection,
square, etc. Being in the environment, the agent has to follow a
well defined direction. The agent objective is to move away by
minimizing interactions with other pedestrians and maximizing its
utility. In other words, the aim of each agent is to walk at the
desired step size and direction, minimizing time travel.

Strategic phase defines the global plan of a pedestrian agent. In
fact, the global plan is the set of step size and moving direction
that the pedestrian has to take in each time interval. This set of
step size and moving direction allows the pedestrian agent to reach
its final goal. At the present, the strategic plan is given to each
pedestrian agent in the beginning of a simulation. In our work, we
focus more on the tactical and the operational phases as follows.

\subsection{Tactical Phase}\label{sec:III_B}
%
%
The tactical phase represents searches for possible positions to
move. These searches are taken at each instant $t$ after the global
plan is set up.
%
The searches taken are based on the position update function, which
is detailed in the following.
%
If the $i^{\rm{th}}$ pedestrian determines moving direction ${\theta
_{i,t}}$, it will try to move a step size ${l_{i,t}}$ in that
direction.
%
%
Due to other pedestrians or obstacles, there could be the case that
movement to the desired position seems to be difficult. For this
kind of cases, step scale factor $\alpha$ and direction shift factor
$\phi$ are needed to be considered.
%
Here, the position update function
%
is formulated as
%
%
\begin{IEEEeqnarray}{lCl}\label{eqn:posupfunc}
f\left( {\alpha ,\phi } \right) = \alpha {l_{i,t}}\left( {\sin
\left( {{\theta _{i,t}} + \phi } \right),\cos \left( {{\theta
_{i,t}} + \phi } \right)} \right){w_{i,t}}
\end{IEEEeqnarray}
%
%
where the walking state factor ${w_{i,t}}$ is defined as
%
%
\begin{IEEEeqnarray}{lCl}\label{eqn:walkstatefactor}
\setlength{\nulldelimiterspace}{0pt} %
{w_{i,t}} = \left\{\begin{IEEEeqnarraybox}[\relax][c]{l's} 1, &with
\,\! $p_{i,t}^{walk}$\\
0, &with \,\! $1 - p_{i,t}^{walk}$ %
\end{IEEEeqnarraybox}\right.
\end{IEEEeqnarray}
%
%
with the walking probability of $0 \le p_{i,t}^{walk} \le 1$.
Through this position update function, the position update process
from time $t$ to $t+1$ can be derived as
%
%
\begin{IEEEeqnarray}{lCl}\label{eqn:posuppro}
{s_{i,t + 1}} = {s_{i,t}} + f\left( {\alpha ,\phi } \right)
\end{IEEEeqnarray}
%
%
where ${s_{i,t}}$ and ${s_{i,t + 1}}$ are the current and the next
positions of the $i^{\rm{th}}$ pedestrian. The corresponding set of
candidate movement position ${S_{i,t + 1}}$ can be expressed as
%
%
%
\begin{IEEEeqnarray}{lCl}\label{eqn:candposset}
{S_{i,t + 1}} = \left\{ {{s_{i,t + 1}} : {\alpha  \in {\rm A},\phi
\in \Phi } } \right\}
\end{IEEEeqnarray}
%
%
where ${\rm A} \!=\! \left\{ {\alpha : {0 \le \alpha  \le 1} }
\right\}$ and $\Phi  \!=\! \left\{ {\phi : { - {\phi _\tau } \le
\phi  \le {\phi _\tau }} } \right\}$. Here we set the constraint
${\phi _\tau } \!<\! \frac{\pi }{2}$ since
%
pedestrians do not choose to move in the opposite direction to the
main crowd flow, even if the direct way subsequently chosen is
crowded \cite{ref:Conf1-38}.
%
%
\begin{figure}[!t]
\centering
\includegraphics[width=0.42\textwidth]{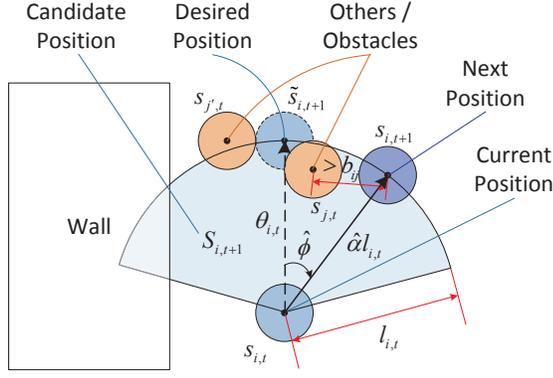}
\caption{Pedestrian movement dynamics.} \label{fig:nextcandpos}
\end{figure}
%
The current position and candidate position set of a pedestrian are
shown in Fig.~\ref{fig:nextcandpos}, for example. In the figure, the
$i^{\rm{th}}$ pedestrian, which is denoted by the blue circle, is
moving in the desired position ${\tilde s_{i,t + 1}}$ from the
current position ${s_{i,t}}$.
%
The candidate position set is depicted as shaded area with sky-blue,
%
%
in which pedestrian can move or remain stationary.
%
The pedestrians in our system are represented by circles because
circles will greatly benefit from geometrical calculation in
programming.

\subsection{Operational Phase}\label{sec:III_C}
%
%
The operational phase describes decisions taken by pedestrian agent
%
to reach the goal
%
at each time $t$ after tactical phase.
%
%
In the operational phase, the pedestrian agent determines the step
size and moving direction suited
%
to avoid people and obstacles.
%
The decisions taken are based on the perception of the environment
where the pedestrian agent is situated. These criteria of decisions
are detailed in the following.
%

%
%
\begin{figure}[!t]
\centering
\includegraphics[width=0.45\textwidth]{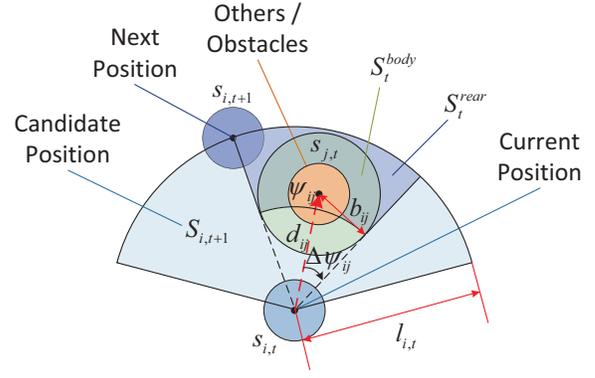}
\caption{Collision avoidance with other members in the crowd.}
\label{fig:CollOther}
\end{figure}
\subsubsection{Collision Avoidance}\label{sec:III_C_1}
In reality, each pedestrian occupies a certain space area. Thus,
avoiding pedestrian overlapping is needed to be considered in
microscopic pedestrian models. By the overlapping, we mean
pedestrians are not allowed to either walk through each other or
share the same space.
%
This assumes that the movement of a pedestrian is affected by other
pedestrians and obstacles in the surrounding environment.



Here, one rule of collision avoidance, which is used to avoid
pedestrian overlapping, is proposed.
%
%
In order to avoid collisions, the positions of other members in the
crowd and those of obstacles should be excluded from the candidate
position set.
%
%
The positions of other people in the collision with the
$i^{\rm{th}}$ pedestrian at time $t$ can be expressed as
%
%
\begin{IEEEeqnarray}{lCl}\label{eqn:set_ped}
S_t^{people} = \left\{ {{s_{j,t}}:\left| {{s_{i,t}} - {s_{j,t}}}
\right| < {l_{i,t}} + {b_{ij}},\forall j \ne i} \right\}
\end{IEEEeqnarray}
%
%
where ${b_{ij}} \!=\! \frac{{{b_i} + {b_j}}}{2}$ from the
$i^{\rm{th}}$ and $j^{\rm{th}}$ pedestrians' body sizes of $b_i$ and
$b_j$, respectively.
%
The body size means the diameter of a pedestrian,
%
%
which makes certain distances among pedestrians and obstacles.
%
The collision positions by other members in the crowd can be
specifically divided into the two parts: body and rear. The
positions in collision with the bodies can be described by
%
%
\begin{IEEEeqnarray}{lCl}\label{eqn:set_body}
S_t^{body} = \left\{ {s:\left| {s - {s_{j,t}}} \right| \le
{b_{ij}},{s_{j,t}} \in S_t^{people}} \right\},
\end{IEEEeqnarray}
%
%
and the positions in the rear of bodies can be explained by
%
%
\newcommand{\sizecorr}[1]{\makebox[0cm]{\phantom{$\displaystyle #1$}}}
\begin{IEEEeqnarray}{lCl}\label{eqn:set_rear}
S_t^{rear} &=& \left\{ {s:\left| {s - {s_{i,t}}} \right| > d_{ij}
{\cos \Delta \psi_{ij} },} \sizecorr{S_t^{people}}
\right.\IEEEnonumber\\
&&\psi_{ij}  - \Delta \psi_{ij}  \le {\rm{atan2}} \left( {\Delta
{x_{i}},\Delta {y_{i}}} \right) \le \psi_{ij} + \Delta \psi_{ij}
, \sizecorr{S_t^{people}} \IEEEnonumber\\
&& \Delta {x_{i}} = x - x_i, \Delta {y_{i}} = y - y_i,
\sizecorr{S_t^{people}} \IEEEnonumber\\
&&\left. {d_{ij} \in D,\psi_{ij} \in \Psi ,\Delta \psi_{ij}  \in
\Delta \Psi } \sizecorr{S_t^{people}} \right\}
\end{IEEEeqnarray}
%
%
where the corresponding sets of distances, directions, and angular
differences are
%
%
\begin{IEEEeqnarray}{lCl}\label{eqn:set_rear}
\!\!D &=& \left\{ {d_{ij}:d_{ij} = \left| {{s_{j,t}} - {s_{i,t}}}
\right|,{s_{j,t}}
\in S_t^{people}} \right\},\\
\!\!\Psi  &=& \left\{ {\psi_{ij} :\psi_{ij}  = {\rm{atan2}} \left(
{\Delta {x_{ji}},\Delta {y_{ji}}} \right)}
\sizecorr{S_t^{people}}, \right. \IEEEnonumber\\
\!\! && \left. \Delta {x_{ji}} = x_j - x_i,\Delta {y_{ji}} = y_j -
y_i, {s_{j,t}} \in S_t^{people} \right\},
\end{IEEEeqnarray}
%
%
and
%
%
\begin{IEEEeqnarray}{lCl}\label{eqn:set_rear}
\!\!\!\!\!\!\!\!\!\Delta \Psi  &=& \left\{ {\Delta \psi_{ij} :\Delta
\psi_{ij} = {{\sin }^{ - 1}}\frac{{b_{ij}}}{d_{ij}},d_{ij} \in D}
\right\},
\end{IEEEeqnarray}
%
%
respectively. The function ${\rm{atan2}} \left( x, y \right)$ is
defined in mathematical term as
\begin{IEEEeqnarray}{lCl}\label{eqn:atan2}
{\rm{atan2}}\left( {x,y} \right) = 2{\tan ^{ - 1}}\left(
{\frac{x}{{\sqrt {{x^2} + {y^2}}  + y}}} \right).
\end{IEEEeqnarray}
The resultant collision positions by other people can be defined as
the union of two sets as $S_t^{other} \!\!=\!\! S_t^{body} \cup
S_t^{rear}$. 
The process of above collision avoidance with other members in the
crowd is elaborated in Fig.~\ref{fig:CollOther}. The collision
positions by obstacles and walls can be derived in a similar manner.

As a result, the positions incurring collisions can be represented
by
%
%
\begin{IEEEeqnarray}{lCl}\label{eqn:colpos}
{S_t^{col}} = S_t^{other} \cup {S^{obstacle}} \cup {S^{wall}},
\end{IEEEeqnarray}
and the candidate position set with collision avoidance is
accomplished through the set difference as
%
%
\setlength{\arraycolsep}{0.7mm}
\begin{IEEEeqnarray}{lCl}\label{eqn:colavoid}
{\hat S_{i,t + 1}} &=& {S_{i,t + 1}}\setminus {S^{col}}
\IEEEyessubnumber\\
&=& \left\{ {s \in {S_{i,t + 1}}\left| {s \notin {S^{col}}} \right.}
\right\} \IEEEyessubnumber\\
&=& \left\{ {{{\hat s}_{i,t + 1}} : {\alpha  \in \hat {\rm A},\phi
\in \hat \Phi } } \right\} \IEEEyessubnumber
\end{IEEEeqnarray}
where ${\hat {\rm A}}$ and ${\hat \Phi }$ are corresponding modified
sets of $\alpha$ and $\phi$, respectively. In this way, the
overlapping positions with other pedestrians and obstacles can be
eliminated in the candidate position set.

\subsubsection{Pedestrian Preference}\label{sec:III_C_2}
To have a realistic simulation of a crowd, we must understand the
behavior of a pedestrian which has preferences for choosing the way
of walking.
%
A pedestrian normally chooses the fastest way to achieve its goal.
In other words, if there are many possibilities, the pedestrian
chooses the straightest way, with the minimum of changing direction,
the most attractive and the less noisy. This is called ``law of
minimal change'' \cite{ref:Conf1-37}. Besides, a pedestrian
typically prefers not to take detours.

Generally, individuals move according to the principle of ``least
effort'' \cite{ref:Conf1-2}. They choose the most familiar and the
easiest one to achieve their goals. They aim to minimize time and
costs by avoiding congestion and by maximizing their step size. If
there is enough time to achieve the goal, a pedestrian chooses to
walk at individually desired step size, corresponding to its most
comfortable walking speed which is called ``least energy-consuming''
\cite{ref:Conf1-2}.

Here, one rule of determining pedestrian step size and moving
direction is proposed, which is used to maximize pedestrian utility.
At each time step $t$, the positions of all pedestrians are updated
synchronously.
%
%
The process of selecting the position to move in the candidate
position set avoiding collision is formulated as
\begin{IEEEeqnarray}{lCl}\label{eqn:nextpos}
{s_{i,t + 1}} = {s_{i,t}} + f\left( {\hat \alpha ,\hat \phi }
\right)
\end{IEEEeqnarray}
%
%
where $\hat \alpha$ and $\hat \phi$ are the step scale factor and
direction shift factor maximizing following utility function,
%
%
%
%
\begin{IEEEeqnarray}{lCl}\label{eqn:utility}
\left( {\hat \alpha ,\hat \phi } \right) = \mathop
{\rm{argmax}}\limits_{\alpha  \in \hat {\rm A},\phi  \in \hat \Phi }
\left[ {{w_\alpha }\alpha  + {w_\phi }\left( {1 - \frac{{\left| \phi
\right|}}{{{\phi _\tau }}}} \right)} \right]
\end{IEEEeqnarray}
with the weighting factors $0 \! \le \! {w_\alpha } \! \le \! 1$ and
$0 \! \le \! {w_\phi } \! \le \! 1$ on step scale and direction
shift, respectively. The weighting factors, ${w_\alpha } + {w_\phi }
= 1$, describe the rate of pedestrian preference on maximizing step
size and minimizing direction change.
%
The resultant next position to move is shown as a circle with dark
blue in Fig.~\ref{fig:nextcandpos} and Fig.~\ref{fig:CollOther}, for
example.
%
%
In this way, the repulsive actions imposed by surroundings will be
strong when a pedestrian closes to other ones or obstacles very much
in movement direction.

Through the collision avoidance and pedestrian preference,
%
%
pedestrians do not overlap each other after the position update
%
%
at
each time interval,
%
although they could move a very small step size, even close to zero.
In other words, their further close at the next time step is
prevented, and the overlapping among them could be avoided.

%
In our model, the maximum movement range of a pedestrian in each
time instant is limited, and hence the other pedestrians and
obstacles in the range only affect the movement position of the
pedestrian.
%
It is very essential to say that in a crowd situation, generally,
the pedestrian is not spontaneous regarding his behavioral strategy.
This is why, the pedestrian tries to have the optimal behaviors
avoiding the collision with the people around and follows the person
in front \cite{ref:Conf1-2}.

%
\section{Simulation Results and Discussion}\label{sec:IV}
We perform Matlab simulations using the proposed model under two
scenarios, i.e., corridor and bottleneck.
%
%
%
In the last decades, by means of experiment or modeling, researchers
have carried out a large number of studies that focus on pedestrian
behavior and movement characteristic, and they have obtained many
landmark achievements.
%
For example, the fundamental diagram and the flow rate at bottleneck
are considered as the main microscopic observables that characterize
pedestrian dynamics in normal state
\cite{ref:SCI2-5, ref:SCI2-4, ref:SCI2-1, ref:SCI2-6, ref:SCI2-7,
ref:SCI2-49, ref:SCI1-10, ref:SCI1-11, ref:SCI2-48, ref:SCI1-3}.
%
To validate the proposed model in this paper, several simulations
are conducted, and then simulation results will be compared with
some classic conclusions and published empirical data provided by
other literature.

\subsection{Fundamental Diagram}\label{sec:IV_A}
%
%
\begin{figure}[!t]
\centering
\includegraphics[width=0.45\textwidth]{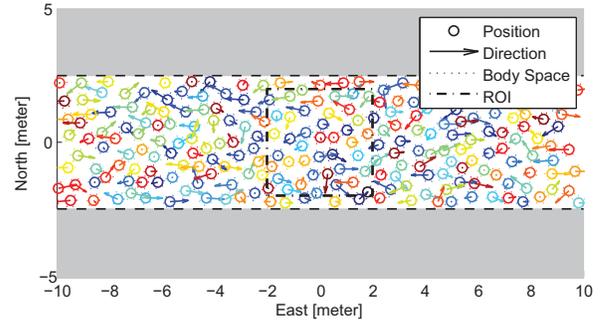}
\caption{Corridor scenario setup.} \label{fig:corridor}
\end{figure}
%
%
\begin{figure}[!t]
\centering
\includegraphics[width=0.45\textwidth]{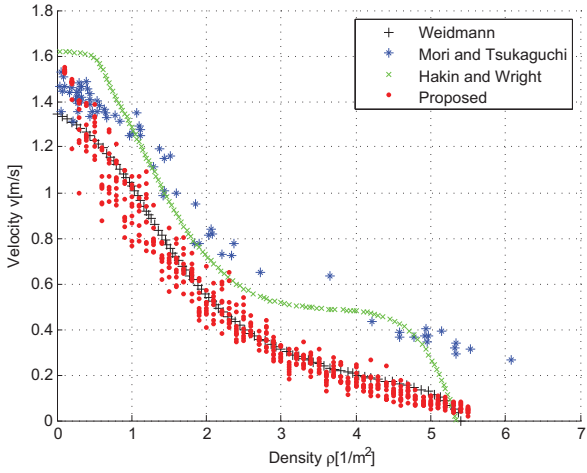}
\caption{Fundamental diagram that shows the density-velocity
relationship.} \label{fig:fundamental}
\end{figure}
%
As shown in Fig.~\ref{fig:corridor}, we conduct a simulation in a
corridor $\left(20\rm{m}\times5\rm{m}\right)$
where each unidirectional pedestrian moves according to the proposed
system.
%
Several studies have investigated the fundamental diagram
%
%
\cite{ref:SCI2-5, ref:SCI2-4, ref:SCI2-1, ref:SCI2-6, ref:SCI2-7,
ref:SCI2-49}
and suggest that the density-speed relation and the density-flow
relation are two of the most important benchmark tests for crowd
simulation. Hence, we validate our model from the viewpoint of the
density-speed relation.
%
%
In general, different density and velocity calculation methods can
be used for pedestrian model \cite{ref:SCI1-12}. These methods are
basically consistent with on another. In this paper, a square area
$\left(2\rm{m}\times2\rm{m}\right)$ is selected as the region of
interest (ROI) to extract the data as in Fig.~\ref{fig:corridor}.
%
In other words, only the data on pedestrian movement in the ROI are
considered.
%
When a pedestrian on one of the two walkways moves in the desired
direction and arrives at the exit boundary, then that pedestrian is
removed from the walkway.
%
At each time instant, the velocity of a pedestrian is calculated by
dividing the step size by the incremental time interval.
%
We iterate simulations ten times at each density and obtain the
density-speed relation within ROI, as shown in
Fig.~\ref{fig:fundamental} (the red dots).
%

Compared with other experimental data of unidirectional flow
\cite{ref:SCI2-5, ref:SCI2-4, ref:SCI2-1}
it is clear that the simulation result has the same tendency with
other density-speed relations. Moreover, the fundamental diagram
obtained from our model is very close to that given by Weidmann
\cite{ref:SCI2-5}.
%
It is indeed difficult to choose a standard fundamental diagram to
evaluate our fundamental diagram. Many researchers have provided the
fundamental diagrams deriving from their own experiments
%
%
where the underlying data of Weidmann's fundamental diagram are
based on the literature research of $25$ publications, including
field studies and experiment research.
%
Compared with other researchers' fundamental diagram, Weidmann's
fundamental diagram may be more comprehensive, and therefore, we
compared our diagram with Weidmann's diagram rather than with other
diagrams.

\subsection{Flow Rate at Bottleneck}\label{sec:IV_B}
%
%
\begin{figure}[!t]
\centering
\includegraphics[width=0.45\textwidth]{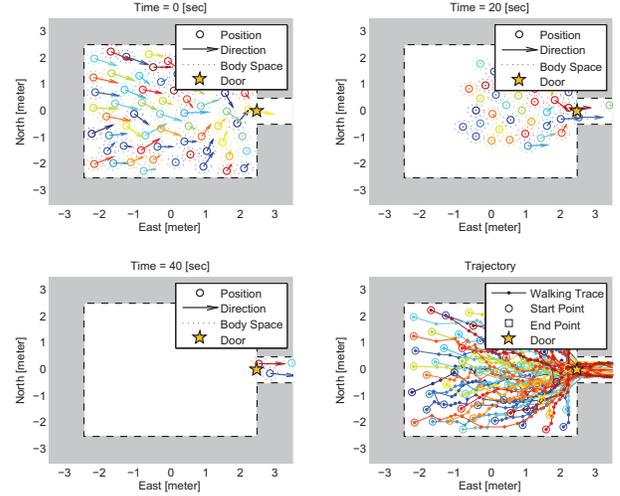}
\caption{Room with door scenario setup.} \label{fig:room}
\end{figure}
%
%
\begin{figure}[!t]
\centering
\includegraphics[width=0.45\textwidth]{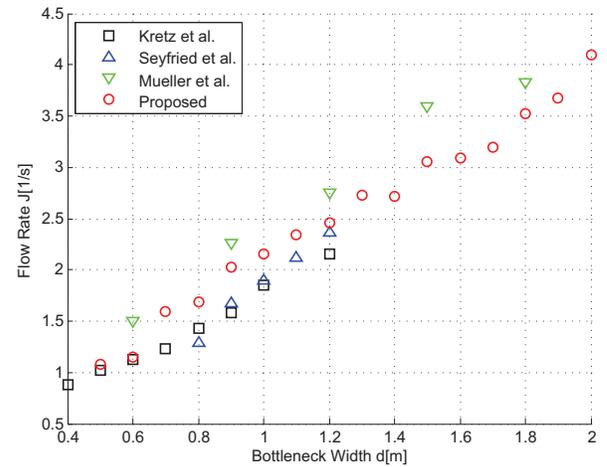}
\caption{Flow rate at bottleneck.} \label{fig:Bottleneck}
\end{figure}
%
%

Flow rate in persons per second or the specific flow rate in persons
per second per meter at the position of bottleneck is one of the
most important parameters for planning and designing facilities,
such as a door.
%
Various published field survey results on flow rate were collected
from different scenarios. Most of these results suggest that flow
rate is nearly in a linear relationship with the width of the
bottleneck.
%
%
In this paper, the pedestrian movement from the room through the
door with changeable width at the right boundary is studied as in
Fig.~\ref{fig:room}. The width of the door is tested with different
values from 0.5m to 2.0m, and the pedestrian density is set to full
capacity to guarantee that the maximum flow rate can be reached at
the door.

The maximum flow rates at bottleneck under different door widths are
%
%
compared with field observation results in Fig.~\ref{fig:Bottleneck}
(the red circles). Three different research results are cited here
for this purpose, including the laboratory observation data
collected by Kretz et al. \cite{ref:SCI1-10}, Seyfried et al.
\cite{ref:SCI1-11, ref:SCI2-48},
and Muller et al. \cite{ref:SCI1-3}.
%
The simulation results show consistency with many empirical data
provided by other literature, which verifies that the proposed model
is applicable to study unidirectional pedestrian model.


\section{Conclusion}\label{sec:V}
Pedestrian crowd is a complex phenomena in which several dangerous
accident can occur, especially in dense situations. An appropriate
solution to this problem is to test the phenomena through the
simulation of pedestrian crowd. This solution could be useful only
if the simulation model produces realistic pedestrian crowd
situations.

For this reason, we present a simple and realistic pedestrian
simulation model which is based on multi-agent systems and that
includes the major human factors.
%
Agents move a step toward the direction calculated by a utility
maximization approach in which various factors that influence
pedestrian movement are considered.
%
%
The system was validated by modeling several basic pedestrian crowd
phenomena and comparing the simulation results with published
empirical data.
%
The conducted simulations show that the model is consistent with the
classic conclusions and published empirical data provided by other
literature, and thus it provides realistic simulated pedestrian
behaviors.
%
%
Here, we have concentrated our work only in the behavior of
normative personality. We can improve our model by considering
different individual personalities.


\section*{Acknowledgment}
This research was done in 2014, but has not been yet published
due to some personal reasons. We decided to release this research
to be opened to public accesss now in 2017.

\bibliographystyle{IEEEtran}
\bibliography{bare_jrnl}



\end{document}